\documentclass[twocolumn]{rbef}
\usepackage{amsmath}
\usepackage{multirow}
\usepackage{array}
\usepackage{caption}
\usepackage{float}
\usepackage{graphicx}
\numeracao{xxxx}
\volume{xx}
\numero{xx}
\ano{2025}
\doi{http://dx.doi.org/xxxx}
\tipodeartigo{Artigos Gerais}
	\author{J. A. S. Lima$^1$} 
	\author{M. H. Benetti$^2$}
	\affil{Universidade de São Paulo, Departamento de Astronomia (IAG-USP) 
		} 

\titulo{Non-Gaussian velocity distributions Maxwell would understand}

\begin{document}
\renewcommand{\tablename}{Table}
\renewcommand{\figurename}{Figure}
\renewcommand{\refname}{References} 
\begin{primeirapagina}

\begin{abstract}
In 1988, Constantino Tsallis proposed an extension of the Boltzmann statistical mechanics by postulating a new entropy formula, $S_q = k_B\ln_q W$, where $W$ is the number of microstates accessible to the system, and $\ln_q$ defines a deformation of the logarithmic function.  This \textit{top-down}\, approach recovers the celebrated Boltzmann entropy in the limit $q\to 1$ since $S_1 = k_B\ln W$. However,  for $q\neq 1$ the entropy is non-additive and has been successfully applied for a variety of phenomena ranging from plasma physics to cosmology. For a system of particles, Tsallis' formula predicts a large class of power-law velocity distributions reducing to the Maxwellian result only for a particular case. Here a more pedagogical \textit{bottom-up} path is adopted. We show that a large set of power-law distributions for an ideal gas in equilibrium at temperature T is derived by slightly modifying the seminal Maxwell approach put forward in 1860. The emergence of power-laws velocity distribution is not necessarily related with the presence of long-range interactions. It also shed some light on the long-standing problem concerning the validity of the zeroth law of thermodynamics in this context. Potentially, since the new method highlights the value of hypotheses in the construction of a basic knowledge, it may have an interesting pedagogical and methodological value for undergraduate and graduate students of physics and related areas.

\end{abstract}
\end{primeirapagina}

\section{Introduction}

It is widely known that Clausius introduced the notion of mean free path for gas particles \cite{RC1858}. However, it was Maxwell who defined the concept of statistical velocity distribution and deduced its functional form for an ideal gas, an essential kinetic result commonly taught in undergraduate courses in physics, chemistry, astronomy, and related fields. In virtue of its methodological importance (and deduction itself), our starting point here is the seminal work published by Maxwell in 1860 entitled \textit{Illustrations of the Dynamical Theory of Gases} \cite{M1860}. 

Maxwell’s velocity distribution initiated a lasting union between the physics of kinetic processes and statistical ideas, thereby promoting the development of classical statistical mechanics. Firstly, through Boltzmann's perspective and later based on a more abstract framework, commonly referred to as statistical ensembles formulated by John Willard Gibbs (1839–1903). In Gibbsian theory, one considers an infinite mental collection of independent systems described by the same Hamiltonian, distributed across different microstates and all consistent with macroscopic constraints \cite{Sbook,Sbook1,Huang,Reif}.

In a point of fact, recent undergraduate (and graduate) textbooks on kinetic theory of gases (\textbf{KTG}) do not present Maxwell’s original derivation. Nevertheless, it will be revisited next section, based only on the core of his hypotheses. Although such a simple proof might be considered somewhat primitive, it is useful from a pedagogical viewpoint in order to show that Maxwell's approach can be slightly modified for deriving  more general  distributions (power-laws). 
       
More recently, much attention has been paid to power-law velocity distributions and their applications to a plethora of phenomena across diverse fields, among them: plasma physics, nuclear physics, astrophysics,  biology, finance, and econometrics, to name just a few.  Among such proposals, the most discussed and applied to date is Tsallis' non-extensive velocity distributions \cite{T88}. 

Tsallis' distribution, also termed $q$-velocity distribution, was first derived based on the theory of statistical ensembles based on Tsallis' $q$-entropic formulation. Some time later, it was also kinetically discussed both in the non-relativistic \cite{SPL98, LSS2000, LSP2001,LSS2002} and relativistic domains \cite{L2002, SL2005, k2010, APS2017}. 

In this context, our aim here is to show that a suitable modification of Maxwell’s original kinetic contribution (i.e. an extension of his first “postulate”), naturally leads to power-law velocity distributions from which the possibility of Gaussian distribution is nothing more than a particular case, already `written' in Maxwell's original hypotheses. 

In principle, a Neo-Maxwellian approach may have two interesting consequences for undergraduate and graduate students. On the one hand, it provides a better understanding of Maxwell's method and also of its generalized version. On the other hand, it may mitigate their curiosity for non-Gaussian velocity distributions and applications in the recent literature\footnote{For a regularly updated bibliography see, for instance, http:tsallis.cat.cbpf.brbiblio.htm}. 

As we shall see, maintaining the appropriate proportions, the confrontation between Maxwellian (Gaussian) and Neo-Maxwellian (Power-Laws) distributions in \textbf{KTG} is similar to the rejection of the fifth postulate of Euclid leading to the emergence of non-Euclidean geometries. In the former case, more general (power-laws) distributions need to be considered, while in the latter, an astonishing window revealed curved spaces in mathematics and, later on, the notion of curved spacetimes in physics. 

The article is planned as follows. Section 2 discusses Maxwell's original derivation, emphasizing what can be used to derive more general distributions. In Section 3, we define the Neo-Maxwellian treatment for describing a large class of non-Gaussian statistics. In Section 4, we will focus on the particular case of Tsallis' distributions, while in Section 5, the basic kinetic predictions of both distributions are confronted (Maxwell versus Tsallis). Finally, in Section 6, our closing remarks and conclusions will be presented.  Details of some calculations are shown in appendices A, B, and C.

\section{Velocity Distribution of an Ideal Gas: Revisiting Maxwell's Original Derivation (1860)}

In his memorable article, Maxwell adopted part of Clausius' model, considering a homogeneous and isotropic gas in the absence of external forces. The nature of the gas and its kinetics were modeled by a large number $ N $ of rigid spheres contained in a volume $ V $, which collide elastically with each other and with the walls of the container. Maxwell introduced the notion of a velocity distribution function, $ F(v_x, v_y, v_z) $, and defined the probability of finding the $ v_x $ component of a gas particle's velocity between $ v_x $ and $ v_x + dv_x $ as $ f(v_x)dv_x $, or more generally.
\begin{equation}\label{E1}
f(v_i)dv_i, \,\,\, i = x,y,z.
\end{equation}
An homogeneous and isotropic gas means that the particle concentration $n$ and other local physical quantities such as temperature and pressure should not depend on the infinitesimal volume element $d^3\textbf{r} = dxdydz$ within the container. Isotropy implies that $ f(v_i) = f(-v_i) $. Furthermore, since there is no preferred direction for the particles in the infinitesimal volume of velocity space, $ d^3\textbf{v} \equiv dv_x dv_y dv_z $, the probabilities must be equal in all directions, not depending specifically on the choice of velocity components. This statistical independence in velocity space can be expressed by the following postulate:

\textbf{PI}: The unknown 3-dimensional velocity distribution is factorizable 
\begin{eqnarray}\label{E2}
F(\textbf{v})d^3\textbf{v} &=& f(v_x,v_y,v_z)dv_x dv_y dv_z\nonumber \\
&=& f(v_x)f(v_y)f(v_z)dv_x dv_y dv_z.
\end{eqnarray}
This postulate has a very simple statistical meaning. The product of probabilities is the mathematical translation of statistical independence. Naturally, in the kinetic case, we refer to the independence of the probability densities of the velocity components, $f(v_i)$.

In addition to the assumed statistical independence, the isotropy in velocity space also suggests the following postulate:

\textbf{PII}: The three-dimensional distribution function $F(\textbf{v})$ depends only on the magnitude of the resulting velocity of each particle, $v = |\textbf{v}|$. This happens because the directions of coordinates are perfectly arbitrary, and, as such, this number must depend on the distance from the origin. 

Mathematically, this postulate can be written as:
\begin{equation}\label{E3}
 F\left(\sqrt{v_x^2 + v_y^2+ v_z^2}\,\right)=f(v_x)f(v_y)f(v_z).
\end{equation}
In reality, Maxwell also considered other propositions connected with the collisions among the hard sphere particles of the gas. Later on, such assumptions were important for his paper of 1867 \cite{M1867} and also for Boltzmann's remarkable work on the H-theorem. Assumptions describing \textbf{PI} and \textbf{PII} are contained in Maxwell's proposition IV \cite{M1860}. Here such assumptions were elevated to the category of basic postulates since they are enough for deriving the Maxwellian distribution and also for understanding how the Neo-Maxwellian extensions arise.  

Let us now derive Maxwell's results.  In order to do that one needs to solve the above functional equations.  For simplicity, let us first determine the 1-Dim probability density $ f(v_x) $ satisfying the above postulates. 

Taking the logarithm of equation (\ref{E3}) and differentiating both sides with respect to $ v_x $, we obtain
\begin{eqnarray}\label{E4}
    \frac{v_x F'}{v\,F} = \frac{d\ln f(v_x)}{dv_x}.
\end{eqnarray}
Repeating this procedure for the other components, we will have
\begin{equation}\label{E5}
    \chi({{v}}) = \frac{1}{v_x}\frac{d \ln f(v_x)}{dv_x} = \frac{1}{v_y}\frac{d \ln f(v_y)}{dv_y} = \frac{1}{v_z}\frac{d \ln f(v_z)}{dv_z},
\end{equation}
where
\begin{equation}\label{E6}
     \chi({v}) = \frac{F'}{vF}. 
\end{equation}

Equation (\ref{E5}) is satisfied only if all terms are equal to a constant which can depend on the mass of the gas particles. Therefore, without loss of generality, we write $ \chi(v) = - \beta m $, where $ \beta$ is some (dimensional) constant characterizing the equilibrium state of the perfect gas. Substituting this relationship into (\ref{E5}) and by analyzing only the $ v_x $ component, we obtain
\begin{equation}\label{E7}
    \frac{1}{v_x}\frac{d\ln f(v_x)}{dv_x}= -\beta m\,\, \Leftrightarrow  \,\,\ln f(v_x) = -\frac{\beta m v_x^2}{2} + \gamma,  
\end{equation}
where $ \gamma $ is another arbitrary constant. By exponentiating both sides of the second expression above and defining $ A_1 = e^{-\gamma} $, it follows that

\begin{equation}\label{E8}
    f(v_x) = A_1\,\exp\left(-\frac{\beta m v_x^2}{2}\right).
\end{equation}
Note that $\text{Dim}|\beta| = \text{Dim}|\text{Energy}|^{-1} $. As we shall see, $\beta= (k_B T)^{-1}$ is determined by comparing the kinetic expression of the pressure in terms of $\beta$ with  the ideal gas equation of state (\textbf{EoS}), a well-established experimental result in classical kinetic theory \cite{Huang}. Surprisingly, we also demonstrate next section that such result remains valid in the Neo-Maxwellian approach.  

The value of the constant $A_1$ is readily obtained defining the total number of particles $N$ contained in the volume $V$ as the integral of the distribution function over the 4-dimensional phase space $dv_xd^3\text{r}$, we have:
\begin{equation}\label{E12}
   N = \displaystyle\int_{-\infty}^{\infty} f(v_x)d v_x\int_Vd^3\,\text{r} = A_1V\left(\frac{2\pi}{\beta m}\right)^{1/2}.
\end{equation}
Note that the above integral on the spatial coordinates yields the volume $V$.  The constant $ A_1 $ follows from the definition of the concentration $ n = N/V$ an also of the well-known Gaussian integral\footnote{\,\,$\int_{0}^{\infty} x^n e^{-\alpha x^{2}} dx = \frac{1}{2}\Gamma \left(\frac{n+1}{2}\right) \alpha^{-(n+1)/2}$\, (see, for instance,  op. cited \cite{Reif} page 609). For $n=0$ we have $\int_{0}^{\infty} e^{-\alpha x^{2}}dx = \frac{1}{2}\sqrt{\frac{\pi}{\alpha}}$.}.

\begin{equation}\label{E9}
    A_1 = n\left(\frac{ \beta m}{2\pi}\right)^{1/2}.
\end{equation}
Substituting $ A_1 $ into (\ref{E8}), we obtain the expression
\begin{equation}\label{E10}
   f(v_x) = n\left(\frac{ \beta m}{2\pi }\right)^{1/2}\exp\left(-\frac{ \beta m v_x^2}{2}\right), 
\end{equation}
representing the (1-D) Maxwell velocity distribution.  

For the 3-D case, it is sufficient to know that the distributions $ f(v_y) $, $ f(v_z) $, have the same functional form of $ f(v_x)$. Hence, one may write: 

 \begin{equation}\label{E11}
 F(v) = f(v_x)f(v_y)f(v_z) = A_3\exp \left(- \frac{\beta m v^2}{2 }\right),
 \end{equation}
 where $A_3$ is the 3-D normalization constant. Defining again the total number of particles $N$ and by integrating in the phase space one finds:

\begin{equation}\label{E13}
A_3 =  n \left(\frac{\beta m}{2\pi}\right)^{3/2}.
\end{equation}
And substituting (\ref{E13}) into (\ref{E11}), the 3-D distribution takes the following form:

\begin{equation}\label{E14}
F(v) = n \left(\frac{\beta m}{2\pi}\right)^{3/2} \exp\left(-\frac{ \beta m v^2}{2 }\right).
\end{equation}
The above Gaussian distribution function is a consequence of the statistical independence contained in the postulates \textbf{PI} and \textbf{PII}. In this strict sense, it should not be considered as particularly surprising.

When we are interested in the probability of finding particle speeds between $v$ and $ v + dv $, there are several combinations of the values of each component yielding in the same speed. Such combinations cover a spherical shell of radius $ v $ and thickness $ dv $. By introducing the speed distribution function
\begin{eqnarray}\label{E15}
\mathcal{F}d^3\textbf{v} &=& F(v)4\pi v^2dv = \mathcal{F}(v)dv  \\
&=&  n\left(\frac{\beta m}{2\pi }\right)^{3/2}4\pi v^2\,\exp\left(-\frac{ \beta m v^2}{2 }\right)dv. \nonumber   
\end{eqnarray}
Given the above distribution (\ref{E15}), we can  calculate the average physical quantities. The average kinetic energy per particle can be written as
\begin{equation}\label{E16}
 \bar{\varepsilon}_M = \frac{\displaystyle\int_{0}^{\infty} \frac{mv^2}{2}\,\,\,F(v)4\pi v^2dv}{\displaystyle\int_{0}^{\infty} F(v)4\pi v^2dv} = \frac{3}{2\beta}.   
\end{equation}
This result is readily obtained by inserting the function $F(v)$ and calculating the Gaussian integrals taking $n=4$ (numerator) and $n=2$ (denominator) in footnote 2.

The kinetic pressure results from collisions between the particles and also with the container walls. It is the average force per unit area exerted by the gas on a reflective surface. For a plane normal to the $ {x} $-direction, only particles with a positive $ p_x $ component contribute. The momentum change per collision is $ 2p_x $, and the number of particles reflected per second is $ v_x F(v) d^3\textbf{v} $. Thus, we can express:
\begin{equation}\label{E17}
P_M = \int_{v_x \geq 0}^{\infty} 2mv_x^2 F(v) \, d^3\textbf{v} =  2\int_{-\infty}^{\infty} \left(\frac{mv_x^2}{2}\right)  F(v) \, d^3\textbf{v}. 
\end{equation}
By symmetry, the average kinetic energy associated with any velocity component is always one-third of the total average kinetic energy. Considering equation (\ref{E16}), this is directly that, 
\begin{equation}\label{E18}
P_M = \frac{2}{3} \int_{0}^{\infty}\frac{mv^2}{2}\, F(v) 4\pi v^2 dv = \frac{2}{3}n\bar{\varepsilon}.
\end{equation}
Thus, the pressure becomes
\begin{equation}\label{E19}
P_M = n\beta^{-1} \,\, \,\, \Leftrightarrow \,\,\,\, P_MV = N\beta^{-1}.
\end{equation}
In order to determine the energy scale $\beta^{-1}$, we recall that the equation of state (\textbf{EoS}) of an ideal gas, consisting of a large number $N$ of particles obeys the empirical relation:
\begin{equation}\label{E20}
P_M V = Nk_BT,
\end{equation}
where $k_B$ is the Boltzmann constant. A comparison of (\ref{E20}) with  (\ref{E19}) reveals that in this limit, $\beta^{-1} = k_B T$. Inserting this relationship into (\ref{E16}) yields the Maxwellian result
\begin{equation}\label{E20b}
\bar{\varepsilon}_{M} = \frac{3k_BT}{2}.
\end{equation}

It is also notable that the parameter $\beta$ can be eliminated by combining equations (\ref{E16}), (\ref{E18}), and (\ref{E19}), yielding a general form for the \textbf{EoS}:
\begin{equation}\label{E20a}
\frac{P_MV}{U_M} = \frac{2}{3},
\end{equation}
where $U_M = N\bar{\varepsilon}_M$ represents the internal energy of the gas. We stress that such \textbf{EOS} is also valid for a bosonic or fermionic quantum ideal gas \cite{LL85}. In the following section, we will demonstrate its validity for distributions described by a power-law.

The calculation of the typical Maxwellian average speeds are widely known (see for instance refs. \cite{Sbook,Sbook1,Huang,Reif}). Hence, we will present here only the basic results:
\begin{eqnarray}\label{E21a}
\begin{aligned}
&\text{\,\,\,\,\,\,\,\,Maxwellian}\\
&\text{typical velocities}
\end{aligned}
\quad
\left\{
\begin{array}{ll}
    \displaystyle v_M^{\text{rms}} = \sqrt{\frac{3k_B T}{m}}, \\[10pt]
    \displaystyle \bar{v}_M = \sqrt{\frac{8k_B T}{\pi m}}, \\[10pt]
    \displaystyle v_M^{\text{mp}} = \sqrt{\frac{2k_B T}{m}}.
\end{array} \right.
\end{eqnarray}
where $v_M^{\text{rms}} $, $ \bar{v}_M $, and $ v_M^{\text{mp}} $ are the root mean square and most probable speeds, respectively. The abbreviation $ M $ indicates that are Maxwellian results. The 3-velocities for fixed values of $m$ and $T$ satisfy the ratios:

\begin{equation}\label{21b}
v_M^{\text{mp}}: \bar{v}_M: v_M^{\text{rms}}  = 1:1.13:1.22.
\end{equation}
In the next section, we will explore a generalization of Maxwell's results. By rejecting the first and consequently part of the second postulate. In this case,  we will obtain a large class of $d$-velocity distributions described by  power-laws. The Maxwellian distribution will be recovered only in the limiting case of statistical independence of its components. Strictly speaking, this limit occurs when the ideal gas contains a large number of particles ($ N \rightarrow \infty $); a state commonly referred to as the thermodynamic limit. However, power-laws describe an ideal gas for a finite number of particles (for a demonstration based on the Liouville theorem and kinetics \cite{LD2020}).

\section{Neo-Maxwellian Approach and Power-Law Distributions}

To begin with, let us consider the Euler exponential function expressed as \cite{AS1972}
\begin{equation}\label{E22}
e^x = \lim_{n \to \infty} \left( 1 + \frac{x}{n} \right)^n.
\end{equation}
A possible generalization of Euler's result consists in defining the \textit{deformation of the exponential function}. Such procedure is performed simply replacing  $ n $ by a continuous free parameter $ d \in \mathcal{R}$, such that  
\begin{equation}\label{E23}
e_{d}(x) =  \left( 1 + dx \right)^{\frac{1}{d}}.   
\end{equation}
The deformation is characterized  by the subscript $d$ in $ e_d(x) $. For each finite value of $ d $, a new function (power-law) will be generated with a different exponent.  In the limit $ d\to 0 $, the function on the left-hand side of (\ref{E23}) recovers the Euler exponential function. 

Now, associated with the deformed exponential, we can also define the deformed logarithm function, $ \ln_d(x) $, as the inverse function of $ e_p(x) $, namely: 
\begin{equation}\label{E25}
    \ln_d(x) = \frac{x^{d} - 1}{d}.
\end{equation}
Given equations (\ref{E23}) and (\ref{E25}), it is easy to deduce the basic property of an inverse function, that is,  $ e_d\left[\ln_d(f)\right] = \ln_d\left[e_d(f)\right] \equiv f $; which is valid for any function $f(x)$.

Henceforth, we will define the velocity distribution using deformed functions while preserving the property of isotropy. However, the postulate \textbf{PI} will no longer be imposed. Consequently, except in the Maxwellian limit, the resulting distribution function will not be factorizable. Postulate \textbf{PI} now takes the form below: 

\textbf{PI*}: The three-dimensional velocity distribution is generally not factorizable, except in the Maxwellian limit ($d\to 0$)

\begin{equation}\label{E26}
F_d(\textbf{v})d^3\textbf{v} = F_d(v_x,v_y,v_z)dv_x dv_y dv_z,  
\end{equation}
where we define the function
\begin{equation}
 F_d(v_x,v_y,v_z) = e_d\left[\ln_df_d(v_x) + \ln_df_d(v_y)+ \ln_df_d(v_z)\right].  
\end{equation}
Notice that taking the limit $ d\to 0 $, $\ln_d f$ goes to $\ln f$ and $e_d (\ln_d f)$ to $f$ thereby recovering the Maxwellian factorization [see (\ref{E2})]. Mathematically, it has also been  written by a slightly different formulation in \cite{SPL98,LSP2001}.

Similarly, and also as a consequence of \textbf{PI*}, the postulate of isotropy \textbf{PII}, is now replaced by: 

\textbf{PII*}: The three-dimensional distribution function $F(\textbf{v})$ depends only on the magnitude of the resulting velocity of each particle, $v = |\textbf{v}|$, but taking into account that the $d$-distributions are not factorizable:
\begin{equation}\label{E26a}
F_d\left(v\right) = e_d\left[\ln_df_d(v_x) + \ln_df_d(v_y)+ \ln_df_d(v_z)\right], 
\end{equation}
where $v=\sqrt{v_x^2 + v_y^2+ v_z^2}$. 

At this point, it is interesting to ask: \textit{What is the form of the velocity distribution function, satisfying the definition (\ref{E26a}), for arbitrary values of $ d $?} 

In order to answer this question, the above functional equation will be solved by the same approach as adopted in the previous section. 

By applying the deformed logarithm on both sides of equation (\ref{E26a}) and differentiating the result  with respect to $ v_x $, we find
\begin{equation}\label{E27}
\frac{v_x}{v}F_d'(v)\frac{d \left[\ln_d F_d(v)\right]}{d F_d(v)} = \frac{d \left[\ln_df_d(v_x)\right] }{d v_x}. 
\end{equation}
Differentiating again with respect to $ v_y $ and $ v_z $, it follows that
\begin{eqnarray}\label{E28}
 \Phi(v) &=& \frac{1}{v_x}\frac{d \left[\ln_df_d(v_x)\right] }{d v_x} = \frac{1}{v_y}\frac{d \left[\ln_df_d(v_y)\right] }{d v_y} \nonumber \\
 &=& \frac{1}{v_z}\frac{d \left[\ln_df_d(v_z)\right] }{d v_y},   
\end{eqnarray}
where we have defined,  
\begin{equation}\label{E29}
\Phi(v) = \frac{F_d'(v)}{v}\frac{d \left[\ln_d F_d(v)\right]}{d F_d(v)}.
\end{equation}
Equation (\ref{E28}) is satisfied only if all terms are equal and constant. Therefore, we will postulate $\Phi(v) = -\beta_d m$ as an extension of the Maxwellian case,  where $\beta_d$ is an arbitrary parameter that may depend on $d$. Thus, taking only the component $v_x$ in equation (\ref{E28}), we have
\begin{equation}\label{E30}
\frac{1}{v_x}\frac{d \left[\ln_df_d(v_x)\right] }{d v_x} = -\beta_d  m. 
\end{equation}
By integrating equation (\ref{E30}), we obtain 
\begin{equation}\label{E31} 
\ln_df_d(v_x) - \ln_dB_1 = -\frac{\beta_d m v_x^2}{2} , 
\end{equation}
here, $\ln_d B_1$ represents an arbitrary constant of integration that has been transferred to the left-hand side of (\ref{E31}) and expressed in this convenient form. It is assumed that $B_1$ depends on the parameter $d$. Using the definition of the deformed logarithm (\ref{E25}), one can straightforwardly derive the following identity:
\begin{equation}\label{E32}
\ln_d f_d(v_x)- \ln_dB_1 = B_1^d\ln_d\left[\frac{f_d(v_x)}{B_1}\right].
\end{equation}
Substituting (\ref{E32}) on the left-hand side of (\ref{E31}) and taking the deformed exponentiation, we obtain the 1-D distribution function:
\begin{equation}\label{E33}
f_d(v_x) = B_1\left[1-\frac{d\beta mv_x^2}{2}\right]^{\frac{1}{d}} \,\,\,\, 
\mbox{with} \,\,\,\, \beta = \beta_d B_1^{-d}.
\end{equation}
By introducing the arbitrary constant $\beta_d$ in (\ref{E30}), the distribution function (\ref{E33}) becomes dependent only on $\beta$. This eliminates the implicit dependence on the deformation parameter in the normalization constant, achieved through the term $\beta_d B_1^{-d}$. This choice does not present any contradictions (see discussion in subsection (3.2))

Similarly to Section 2, we will only indicate the 1-D functions, $F_d(v_x)$, for $d>0$ and $d<0$, as the calculations are analogous to the 3-D case, which will be presented in Section (3.1).

For the condition $d>0$, we obtain the normalized power low distribution function as follows
\begin{equation}\label{E34}
f_d(v_x) = n \left(\frac{\beta m}{2\pi }\right)^{1/2}  \frac{ d^{1/2} \,\,\Gamma\left(\frac{1}{d} + \frac{3}{2}\right)}{ \Gamma\left(\frac{1}{d} + 1\right)} \left[1 - \frac{ d\beta m v_x^2}{2 }\right]^{\frac{1}{d}},
\end{equation}
where in the integration we made use of the \textit{Beta function}\footnote{see identity $B(x,y) =  \int_{0}^{1} t^{x-1}(1-t)^{y-1}dt = \displaystyle\frac{\Gamma(x)\Gamma(y)}{\Gamma(x+y)}$ in \cite{AS1972}.\\ }.

\begin{figure}[h]
    \centering
    \includegraphics[width=0.5\textwidth]{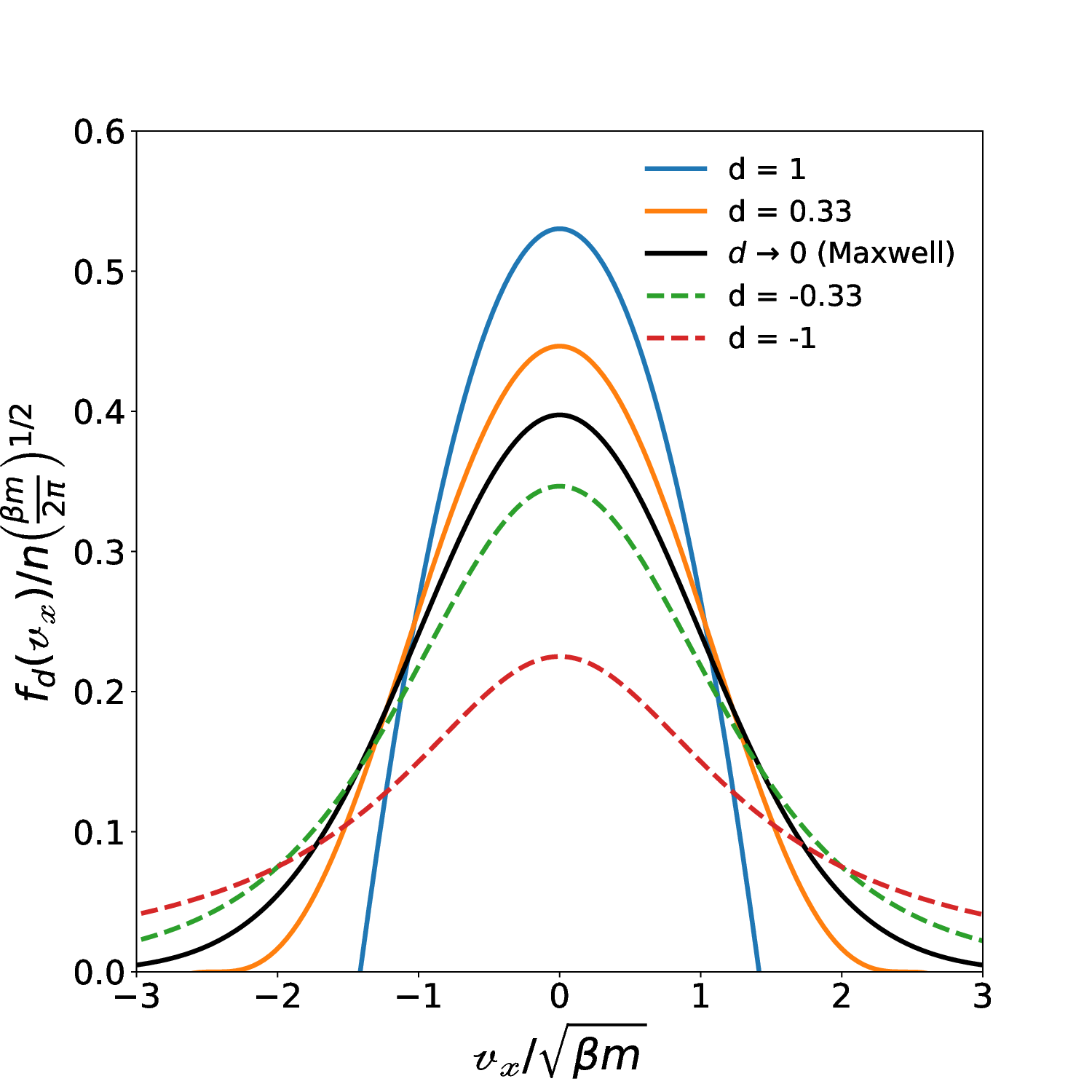}
    \caption{Power-law normalized distributions (1-D). The solid black line represents the Maxwellian limit ($d\to 0$), while the solid lines with $d>0$ tend to zero below the Maxwellian (short-tail). The dashed lines refer to distributions with $d<0$ and decay above the Maxwellian (fat-tail).} 
    \label{f1}
\end{figure}

Now, in the case  $ d<0 \, (d = -|d|)$, the power-law distribution function becomes\footnote{see identity   $B(x,y) =  \int_{0}^{\infty} t^{x-1}(1+t)^{-(y+x)}dt = \displaystyle\frac{\Gamma(x)\Gamma(y)}{\Gamma(x+y)}$ in \cite{AS1972}}
\begin{equation}\label{E35}
f_d(v_x) = n  \left(\frac{ \beta m}{2\pi }\right)^{1/2} \frac{|d|^{1/2}\Gamma\left(\frac{1}{|d|}\right)}{\Gamma\left(\frac{1}{|d|} - \frac{1}{2}\right)} \left[1+\frac{|d| \beta m v_ x^2}{2 }\right]^{-\frac{1}{|d|}}.   
\end{equation}
In \textbf{Figure} \ref{f1}, we show the power-laws for some selected values of the deformation parameter 
$d$. Both the function and its argument were divided by convenient factors, reducing the axes dimensionless. The solid black line represents the $d\to 0$ limit, where the classical Gaussian Maxwell result is recovered. The colored solid lines indicate the power-laws ($d>0$) described by equation (\ref{E34}), known as \textit{short-tail} distributions, as they tend to zero below the Gaussian. The dashed lines represent the power-laws ($d<0$) according to (\ref{E35}), known as \textit{fat-tail} distributions, as they tend to zero above the Gaussian.

In conclusion, from the distributions (\ref{E34}) and (\ref{E35}), we can determine the average energy per particle in one dimension, such that
\begin{equation}\label{E36}
\bar{\varepsilon} = \left\{
\begin{array}{ll}
 \displaystyle\frac{1}{2\beta} \left( \frac{2}{2 + 3d} \right), & \,\, d>0, \\[10pt]
 \displaystyle\frac{1}{2\beta} \left( \frac{2}{2 - 3|d|} \right), & \,\, |d| < \frac{2}{3}.
\end{array}
\right.
\end{equation} 
The second equation in (\ref{E36}) implies the constraint $ |d| < 2/3 $ with $ d<0 $; in other words, we must consider the validity range $ -2/3 < d<0 $ to avoid negative energy.

In the next section, we will show in detail how to determine the 3-D function and the average values.

\subsection{Power-laws in the 3-dimensional case}

It is simple to extend the distributions (\ref{E34}) and (\ref{E35}) to higher dimensions. As an example, we will derive the generalization in three dimensions. In this case, taking the deformed logarithm from equation (\ref{E26a}) and using (\ref{E31}), we obtain 
\begin{equation}\label{E38}
    \ln_d F_d(v) - \ln_d B_{3} = -\frac{\beta_d m v_x^2}{2} - \frac{\beta_d m v_y^2}{2} -\frac{\beta_d m v_z^2}{2},
\end{equation}
where $v^2 = v_x^2+v_y^2+v_z^2$. Now we define the normalization 3-D constant $ B_{3} $. By rewriting the left-hand side of equation (\ref{E38}) using the identity (\ref{E32}) and subsequently applying the deformed exponential, we derive the 3-D distribution function.
\begin{equation}\label{E39}
    F_d(v) = B_3\left[1-\frac{d\beta mv^2}{2}\right]^{\frac{1}{d}} \quad \mbox{com}\quad \beta = \beta_dB_3^{-d}.
\end{equation}
Condition $d>0$ in equation (\ref{E39}) requires the distribution function to be limited by a maximum velocity in order to avoid negative values. However, when $ d<0 $, the function will not be limited, and the integration is carried out over the interval $ (-\infty, +\infty) $. Let us see how to obtain (for all values of d) the normalization constant and other quantities of interest, such as: \textit{average kinetic energy, pressure, root mean square speed, mean speed and the most probable speed}.

\subsection{Case $d>0$}

As mentioned, in this situation, the distribution function (\ref{E39}) will be limited by a maximum velocity, $v_{max}$, which prevents it from generally becoming complex for velocities greater than $v_{max}$. Note that distribution (\ref{E39}) must satisfy the constraint for each component
\begin{equation}\label{E40}
 v_{max} \leq \sqrt{\frac{2 }{d\beta m}}.  
\end{equation}
The normalization constant $ B_3 $ is calculated using equation (\ref{E39}), by solving the following integral
\begin{equation}\label{E41}
n=\int\int\int_{-v_{max}}^{v_{\text{max}}} B_3 \left[1 - \frac{d\beta m v^2}{2 }\right]^{\frac{1}{d}} d^3 \textbf{v},
\end{equation}
where we implicitly consider the volume $ V $, which results from the integral in spatial coordinates, writing $ N/V = n $ we can solve the above integral (see details in the Appendix \ref{A})
\begin{equation}\label{E45}
B_3 = n \left(\frac{ \beta m }{2\pi }\right)^{3/2} \,\,\,\displaystyle\frac{d^{3/2} \Gamma\left(\frac{1}{d}+\frac{5}{2}\right)}{\Gamma\left( \frac{1}{d} +1 \right)},
\end{equation}
where the identity $\Gamma(3/2) = \sqrt{\pi}/2$ was used. Finally, substituting (\ref{E45}) into (\ref{E39}), we find the 3-D power-law  
\begin{equation}\label{E46}
F_d(v) = n \left(\frac{ \beta m }{2\pi }\right)^{3/2} \,\,\,\displaystyle\frac{d^{3/2} \Gamma\left(\frac{1}{d}+\frac{5}{2}\right)}{\Gamma\left( \frac{1}{d} +1 \right)} 
\left[1-\frac{d\beta mv^2}{2 }\right]^{\frac{1}{d}}.   
\end{equation}
From (\ref{E15}) and (\ref{E46}) one can compute the 3-D distribution for speed
\begin{eqnarray} \label{E47}
 \mathcal{F}_p(v) &=& n \left(\frac{\beta m }{2\pi }\right)^{3/2}4\pi v^2 \,\,\,\displaystyle\frac{d^{3/2} \Gamma\left(\frac{1}{d}+\frac{5}{2}\right)}{\Gamma\left( \frac{1}{d} +1 \right)} 
\nonumber \\
&\times& \left[1-\frac{d\beta mv^2}{2 }\right]^{\frac{1}{d}}.   
\end{eqnarray}
In the limit $ d\to 0 $ it can be verified\footnote{see ref. \cite{AS1972},$ \,\,\displaystyle\lim_{|z| \to \infty} \frac{z^{b-a} \Gamma(z+a)}{\Gamma(z+b)} = 1 $.} that equation (\ref{E47}) recovers the Maxwellian case (\ref{E15}), because
\begin{equation} \label{E48}
\left(\frac{d^{3/2} \Gamma\left(\frac{1}{d}+\frac{5}{2}\right)}{\Gamma\left( \frac{1}{d} +1 \right)}\right)_{d\to0} = 1, \quad B_{3} \to n \left( \frac{ \beta m}{2\pi }\right)^{3/2}.
\end{equation}

To calculate the average kinetic energy per particle, we integrate the function (\ref{E47}), which yields
\begin{equation}\label{E49}
\bar{\varepsilon}_d = \frac{B_3\displaystyle\int_{0}^{v_{max}} \left[1-\frac{d\beta mv^2}{2 }\right]^{\frac{1}{d}} \frac{mv^2}{2}\,\,\, 4\pi v^2 dv}{B_3\displaystyle\int_{0}^{v_{max}} \left[1-\frac{d\beta mv^2}{2 }\right]^{\frac{1}{d}}4\pi v^2dv}. 
\end{equation}
Writing equation (\ref{E49}) in terms of $u = dmv^2 / 2 k_B T$ and  using the \textit{Beta function} (see footnote 3), with the help of the recurrence formula $z\Gamma(z) = \Gamma(z+1)$ \cite{AS1972}, we obtain (see Appendix \ref{B}) 

\begin{equation}\label{E52}
\bar{\varepsilon}_d = \frac{3 }{2 \beta} \, \left(\frac{2}{2+5d}\right).
\end{equation}

Now, from equation (\ref{E52}) and following the approach in (\ref{E17}), the pressure is given by \cite{Huang}
\begin{eqnarray}\label{E54}
P_d &=&  \int_{v_x \geq 0}^{v_{max}} 2mv_x^2 F_d(v) \, d^3\textbf{v}  \nonumber \\
&=& \frac{2}{3}  B_3  \int_{0}^{v_{max}}  \left( 1- \frac{d\beta m v^2}{2}\right)^{\frac{1}{d}} \frac{mv^2}{2}\,\,\, 4\pi v^2 dv. \nonumber \\
\end{eqnarray}
It can be verified, as in the Maxwellian case, the equation above is identical to the mean kinetic energy (\ref{E49}) multiplied by $2n/3$. Therefore, by inserting (\ref{E49}) in (\ref{E54}), we conclude that
\begin{equation}\label{E55}
P_d =  \frac{2}{3}n\bar{\varepsilon} = \frac{n}{\beta} \left(\frac{2}{2+5d}\right).
\end{equation}
As expected, equations (\ref{E52}) and (\ref{E55}) recover the Maxwellian results (\ref{E16}) and (\ref{E17}) in the limit $d\to 0$. Following the procedure in Section 2, we can also eliminate $\beta$. Defining $U_d \equiv N\bar{\varepsilon}_d$,  we obtain an equation of state independent of $d$ and $\beta$, exactly in the form of (\ref{E20a}):
\begin{equation}\label{E56}
    \frac{P_d V}{U_d} = \frac{2}{3}.
\end{equation}
a result previously obtained from a kinetic approach within Tsallis' $q$-statistics \cite{LS2005}.  

It is worth notice that the integration method enabled us to detach the standard energy scale $\beta^{-1}$ of the parameter $d$. For consistency, we see that the same result holds if we consider the average energy per particle ($ \varepsilon $) in (\ref{E52}). However, in thermodynamics one may choose 2 independent variables, for instance, the pair ($T,n$). For a fixed V, one may take $N$ as the second variable, which suggests that the power index $d$ may be a function of N, the total number of particles. Since the Maxwellian distribution is recovered when $d\to 0$ and the thermodynamic limit arrives just when $N \to \infty$, a plausible relation would be $d \propto N^{-\delta}$, where delta is a positive number to be determined. As we shall see in the following, such results may cure some difficulties plaguing the Tsallis statistics, such as, for instance, the validity of the zeroth law of thermodynamics (see also discussion in Section 4).    

For completeness and future comparison with the Gaussian case, we present the typical velocities within the 3-D power-law formalism (see Appendix \ref{B}):
\begin{equation}\label{E56a}
\begin{aligned}
&\text{Typical velocities} \\
&\text{\,\,\,\,of power-laws}\\
&\text{\,\,\,\,\,\,\,\,\,\,\,\,\,\,\,$(d>0)$}
\end{aligned}
\left\{
\begin{aligned}
   &v_{d}^{\text{rms}}  = \sqrt{\frac{3k_BT}{ m} \left(\frac{2}{2+5d}\right)}, \\[4pt]
   &\bar{v}_d = \sqrt{\frac{8 k_BT}{\pi m}} \frac{\Gamma\left(\frac{1}{d}+ \frac{5}{2}\right)}{d^{1/2}\Gamma\left(\frac{1}{d} +3\right)}, \\[4pt]
   &v_d^{\text{mp}} = \sqrt{\frac{2k_BT}{ m (1 + d)}}.
\end{aligned}
\right.
\end{equation}
As expected, $v_d^{\text{rms}}$, $\bar{v}_d$, and $v_d^{\text{mp}}$ recover the Maxwellian results (\ref{E21a}) in the limit $d\to 0$.

\subsection{Case $d<0$}

Under this condition, the distribution function can be rewritten using the relation $ |d| = -d $ as follows
\begin{equation}\label{E61}
F_d(v) = B_3 \left[1 + \frac{|d|  m v^2}{2k_BT}\right]^{-\frac{1}{|d|}}.
\end{equation}
The sign change inside the brackets in equation (\ref{E61}) implies that the function is no longer limited by a maximum velocity. In this situation, we find that the normalization constant takes the following form
\begin{equation}\label{E62}
B_3 = n\left(\frac{ m}{2\pi k_BT}\right)^{3/2} \frac{|d|^{3/2} \Gamma\left(\frac{1}{|d|}\right)}{ \Gamma\left(\frac{1}{|d|} - \frac{3}{2}\right)}, 
\end{equation}
Here we use the \textit{Beta function} (see footnote 4). Substituting the normalization (\ref{E62}) into (\ref{E61}), we arrive at the 3-D distribution function 
\begin{equation}\label{E63}
F_d(v) = n\left(\frac{  m}{2\pi k_BT}\right)^{3/2} \frac{|d|^{3/2} \Gamma\left(\frac{1}{|d|}\right)}{ \Gamma\left(\frac{1}{|d|} - \frac{3}{2}\right)} 
\left[1 + \frac{|d|  m v^2}{2k_BT}\right]^{-\frac{1}{|d|}}.       
\end{equation}
Using (\ref{E15}), the 3-D distribution for speed becomes
\begin{eqnarray} \label{E64}
\mathcal{F}_d(v) &=& n\left(\frac{  m}{2\pi k_BT}\right)^{3/2} 4\pi v^2 \frac{|d|^{3/2} \Gamma\left(\frac{1}{|d|}\right)}{ \Gamma\left(\frac{1}{|d|} - \frac{3}{2}\right)} \nonumber \\ 
&\times& \left[1 + \frac{|d|  m v^2}{2k_BT}\right]^{-\frac{1}{|d|}}.    
\end{eqnarray}

The standard classical function (\ref{E15}) is easily recovered for $ d\to 0 $.

Next, we have the average kinetic energy per particle in the following way
\begin{equation}\label{E65}
\bar{\varepsilon}_d = \frac{3k_BT}{2}\left(\frac{2}{2-5|d|}\right), \quad |d| < \frac{2}{5}.
\end{equation}
\begin{figure}[h]
    \centering
    \includegraphics[width=0.5\textwidth]{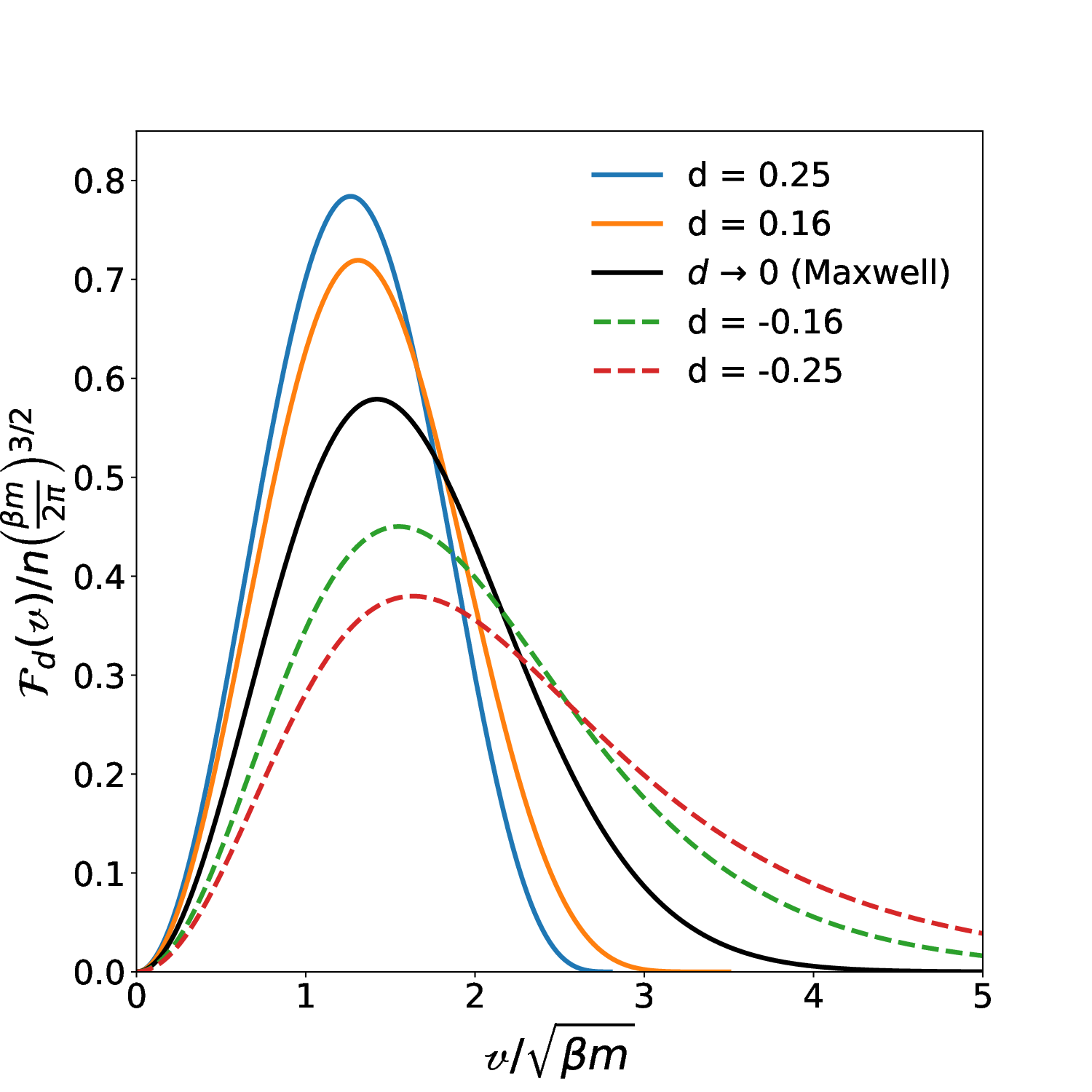}
    \caption{Power-law normalized speed distributions (3-D). The solid black line represents the Maxwellian limit ($ d\to 0 $), while the solid colored lines with $ d>0 $ tend to zero on the right side, below the Maxwellian (short-tail), with peaks shifted to the left. The dashed lines represent the distributions with $ d<0 $, which tend to zero above the Maxwellian (fat-tail), with peaks shifted to the right.}
    \label{f2}
\end{figure}
The constraint $ |d| < 2/5 $ when $ d<0 $, which can be written as $ -2/5 < d < 0 $, is added to avoid negative values of energy.  
Consequently, from (\ref{E65}), the pressure becomes
\begin{equation}\label{E66}
P_d = nk_BT\left(\frac{2}{2-5|d|}\right).
\end{equation}
Both equations (\ref{E65}) and (\ref{E66}) recover the classical case when $ |d| \to 0 $.

Finally, the root mean square speed, the average speed, and the most probable speed will be, respectively:
\begin{equation}\label{E67}
\begin{aligned}
&\text{Typical velocities} \\
&\text{\,\,\,\,of power-laws}\\
&\text{\,\,\,\,\,\,\,\,\,\,\,\,\,\,\,\,\,$(d<0)$}
\end{aligned}
\left\{
\begin{aligned}
    &v_d^{\text{rms}} = \sqrt{\frac{3k_BT}{ m} \left(\frac{2}{2-5|d|}\right)},  \\[4pt]
    &\bar{v}_d = \sqrt{\frac{8 k_BT}{\pi m}} \frac{\Gamma\left(\frac{1}{|d|}- 2\right)}{|d|^{1/2}\Gamma\left(\frac{1}{|d|} -\frac{3}{2}\right)},  \\[4pt]
    &v_d^{\text{mp}} = \sqrt{\frac{2k_BT}{ m (1 - |d|)}}.
\end{aligned} 
\right.
\end{equation}

\begin{table*}[h!]
\centering
\renewcommand{\arraystretch}{2.8} 
\caption{Tsallis distribution functions.}
\label{T2}
\begin{minipage}[t]{0.9\textwidth}
\[
\begin{array}{>{\centering\arraybackslash}m{1cm} c |c}

\hline \hline
\multicolumn{2}{c|}{\textbf{Cases}} & \textbf{Tsallis Distribution} \\ \hline
\multirow{2}{*}{\raisebox{-1.5ex}{\rotatebox{90}{\textbf{Short-Tail}}}} & \text{1-D ($q>1$)} & 
\displaystyle n\left(\frac{m}{2\pi k_BT}\right)^{1/2} \frac{(q-1)^{1/2} \Gamma\left(\frac{1}{q-1} + \frac{3}{2}\right)}{\Gamma\left(\frac{1}{q-1} + 1\right)} 
\left[1 - (q-1)\frac{m v_x^2}{2k_BT}\right]^{\frac{1}{q-1}} \\ 
 & \text{3-D ($q>1$)} & 
\displaystyle n\left(\frac{m}{2\pi k_BT}\right)^{3/2} \frac{(q-1)^{3/2} \Gamma\left(\frac{1}{q-1} + \frac{5}{2}\right)}{\Gamma\left(\frac{1}{q-1} + 1\right)} 
\left[1 - (q-1)\frac{m v^2}{2k_BT}\right]^{\frac{1}{q-1}} \\ 

\multirow{2}{*}{\rotatebox{90}{\textbf{Fat-Tail}}} & \text{1-D ($1/3 < q<1$)} & 
\displaystyle n\left(\frac{m}{2\pi k_BT}\right)^{1/2} \frac{(1-q)^{1/2} \Gamma\left(\frac{1}{1-q}\right)}{\Gamma\left(\frac{1}{1-q} - \frac{1}{2}\right)} 
\left[1 + (1-q)\frac{m v_x^2}{2k_BT}\right]^{-\frac{1}{1-q}} \\ 
 & \text{3-D ($3/5 < q<1$)} & 
\displaystyle n\left(\frac{m}{2\pi k_BT}\right)^{3/2} \frac{(1-q)^{3/2} \Gamma\left(\frac{1}{1-q}\right)}{\Gamma\left(\frac{1}{1-q} - \frac{3}{2}\right)} 
\left[1 + (1-q)\frac{m v^2}{2k_BT}\right]^{-\frac{1}{1-q}} \\ \hline \hline

\end{array}
\]
\end{minipage}
\end{table*}
In \textbf{Figure} \ref{f2}, we show the results of the speed distribution functions (\ref{E47}) and (\ref{E64}) in the dimensionless form. Solid colored lines represent the short-tailed power-laws ($d>0$), as $d\to 0$ their tails approach the Maxwellian from within. Additionally, we observe that the peaks are shifted to the left and decay on top of the Maxwellian. The dashed colored lines represent the fat-tail power-laws ($d<0$), as $d\to 0$ their tails approach the Maxwellian from outside. Moreover, we see that their peaks are shifted to the right and decay under the Maxwellian.

\section{Case Study: Tsallis Power-Laws}

Tsallis' power-law distributions \cite{T88} are based on the non-extensive formulation of statistical mechanics quantified by a positive index $q$. Non-extensive refers to the generalization of Boltzmann-Gibbs (BG) statistical mechanics. For $q \to 1$, the Tsallis entropy $S_q$ reduces to the classical entropy $S_{BG}$, which is said to be \textit{additive}, meaning that for any two independent subsystems, the entropy $S_{BG}$ of their sum coincides with the sum of their individual entropies. In the case $q \neq 1$, the entropy $S_q$ violates this property and is referred to as \textit{non-additive} (sub-additive for $q<1$ and super-additive if $q<1$). It has also been proved that Tsallis statistics provides a suitable theoretical framework for analysing complex systems \cite{TsallisB}. In particular, it has been applied since long ago for several systems in plasma physics \cite{LSS2000,SAL2005}, high energy physics \cite{Cleymans12,CW2}, stellar and galactic processes \cite{LS2005,LR2005} and thermofractals \cite{AD15,MLD2022}.  

The procedure for obtaining the Tsallis distribution from the velocity $d$-distribution function is straightforward. By setting $d = q-1$, the case $ q>1 $ yields the results obtained for $ d>0 $, which correspond to the distribution constrained by the maximum velocity. When $ d<0 \,\, (\text{we use } d = -|d|) $, the relation $ d = q-1 $ can be rewritten as $ |d| = 1-q $ (note that we are not changing the mapping, as $ -|d| = q-1 = d $). In the latter case, we must consider the restriction $ 1/3 < q<1 $, imposed by (\ref{E36}), for 1-D case and $ 3/5 < q < 1 $, as given by (\ref{E65}), for the 3-D case, this latter constraint was also independently derived, through a variational method, in the context of nonlinear wave propagation in plasmas \cite{silveira21a}.

\textbf{Table} \ref{T2} displays different scenarios for Tsallis' distribution function. These results must be compared with the expressions previously derived by Lima \& Silva \cite{LS2005} under the transformation $q^* = 2 - q$. As expected, in the limit $ q\to 1 $ both 1-D and 3-D functions recover the Maxwell distributions (\ref{E10}) and (\ref{E14}), respectively. 

At this point, it is interesting to discuss two intertwined questions about Tsallis statistics: \textit{(i) What about the value of $q$ and the thermodynamic limit in this context?, and (ii) Is the zeroth law of thermodynamics valid for power-law distributions?} 

The first question was resolved a couple of years ago based on the Liouville Theorem. For short tails, it was demonstrated that $q = 1 + \frac{2}{3N}$, where $N$ is the total number of particles. In general $q= 1 \pm 2/3N$ . This means that $d = q-1 = \pm 2/3N$, thus fixing $\delta = 1$ \cite{LD2020,LB2025}. Naturally, other values are possible for different classes of systems (see the comment below equation (\ref{E56}).  

As expected, in the limit $N \to \infty$, $q \to 1$, so that $d = q - 1 \to 0$. Hence, only in the thermodynamic limit the Gaussian distribution is recovered. In particular, we see that replacing the Gaussian postulate \textbf{PI} by the non-Gaussian \text{PI*} means that out of the thermodynamic limit, all power-laws describe an ideal gas, at temperature T, with a \textbf{finite number of particles}.

The second question is more delicate. The validity of the zeroth law motivated a long-standing controversy related to the definition of temperature for thermal equilibrium states in the context of Tsallis' statistics \cite{Nauenberg2003,Tsallis2004,Nauenberg2004}. However, the temperature in our approach is well defined and does not depend on the value of $d=d(N)=q(N) - 1$. Thus, the zeroth law of thermodynamics is protected for all the values of the $q$-parameter. The Neo-Maxwellian approach provides an interesting thermal example, where all power-law distributions possess the same temperature and the zeroth law of thermodynamics is naturally satisfied.

On the other hand, as assumed by many authors working with Tsallis' statistics, one may also think that the $q$-parameter might be a consequence of long-range forces so that the thermal equilibrium condition is not satisfied, and the zeroth law invalidated. As firstly shown by Tolman and Tolman and Ehrenfest \cite{Tolman1930,TE1930}, this happens, for instance, in the presence of a static gravitational field (for a recent discussion, see \cite{Lima2019}). However, clearly this is not the case discussed in the present paper.  Of course, at least in principle, this is not forbidden  for  different contexts.

\section{Maxwell versus Tsallis: Physical Quantities}

\begin{table*}[h!]
\centering
\renewcommand{\arraystretch}{2.8} 
\caption{Comparison of basic results (3-D case). }
\label{T1}
\begin{minipage}[t]{1\textwidth}
\[
\begin{array}{c c c c}

\hline \hline 
\multirow{2}{*}{\text{}} & 
\multicolumn{1}{c|}{\parbox[c]{2.5cm}{\centering \textbf{Gaussian (Maxwellian)} \\ $q \to 1$}} & 
\multicolumn{1}{c|}{\parbox[c]{2.5cm}{\centering \textbf{power-law (Short-Tail)} \\ $q>1$}} & 
\multicolumn{1}{c}{\parbox[c]{2.5cm}{\centering \textbf{power-law (Fat-Tail)} \\ $3/5 < q<1$}} \\ 
\hline \hline
\text{$\bar{\varepsilon}$} &  \displaystyle\frac{3k_BT}{2} & \displaystyle\frac{3k_BT}{2}\left(\frac{2}{5q-3}\right) &  \displaystyle\frac{3k_BT}{2}\left(\frac{2}{5q-3}\right) \\ 
\text{$P$} &  \displaystyle nk_BT &  \displaystyle nk_BT\left(\frac{2}{5q-3}\right) &  \displaystyle nk_BT\left(\frac{2}{5q-3}\right) \\ 
\text{$v^\text{rms}$} &  \displaystyle\sqrt{\frac{3k_BT}{ m} } & \displaystyle\sqrt{\frac{3k_BT}{ m} \left(\frac{2}{5q-3}\right)} &  \displaystyle\sqrt{\frac{3k_BT}{ m} \left(\frac{2}{5q-3}\right)} \\ 
\text{$\bar{v}$} &   \displaystyle\sqrt{\frac{8k_BT}{ \pi m }} &\displaystyle\sqrt{\frac{8k_BT}{ \pi m }}\frac{\Gamma\left(\frac{1}{q-1} + \frac{5}{2}\right)}{ (q-1)^{\frac{1}{2}}\Gamma\left(\frac{1}{q-1}+3\right)}   & \displaystyle\displaystyle\sqrt{\frac{8k_BT}{ \pi m }}\frac{\Gamma\left(\frac{1}{1-q}- 2\right)}{(1-q)^{\frac{1}{2}}\Gamma\left(\frac{1}{1-q} -\frac{3}{2}\right)}  \\ 
\text{$v^{\text{mp}}$} &   \displaystyle\sqrt{\frac{2k_BT}{ m }} &   \displaystyle\sqrt{\frac{2k_BT}{ m q}} &  \displaystyle\sqrt{\frac{2k_BT}{ m q}}   \\ 
\hline \hline
\end{array}
\]
\end{minipage}
\end{table*}

To compare the basic results derived from Tsallis' distributions, we use equations (\ref{E52}), (\ref{E55}), and (\ref{E56a}) for the short-tail physical quantities, and equations (\ref{E65}), (\ref{E66}), and (\ref{E67}) for the fat-tail physical quantities. The results including those for the Maxwellian case are summarized in \textbf{Table} \ref{T1}. 
\begin{figure}[h!]
\centering
\includegraphics[width=0.5\textwidth]{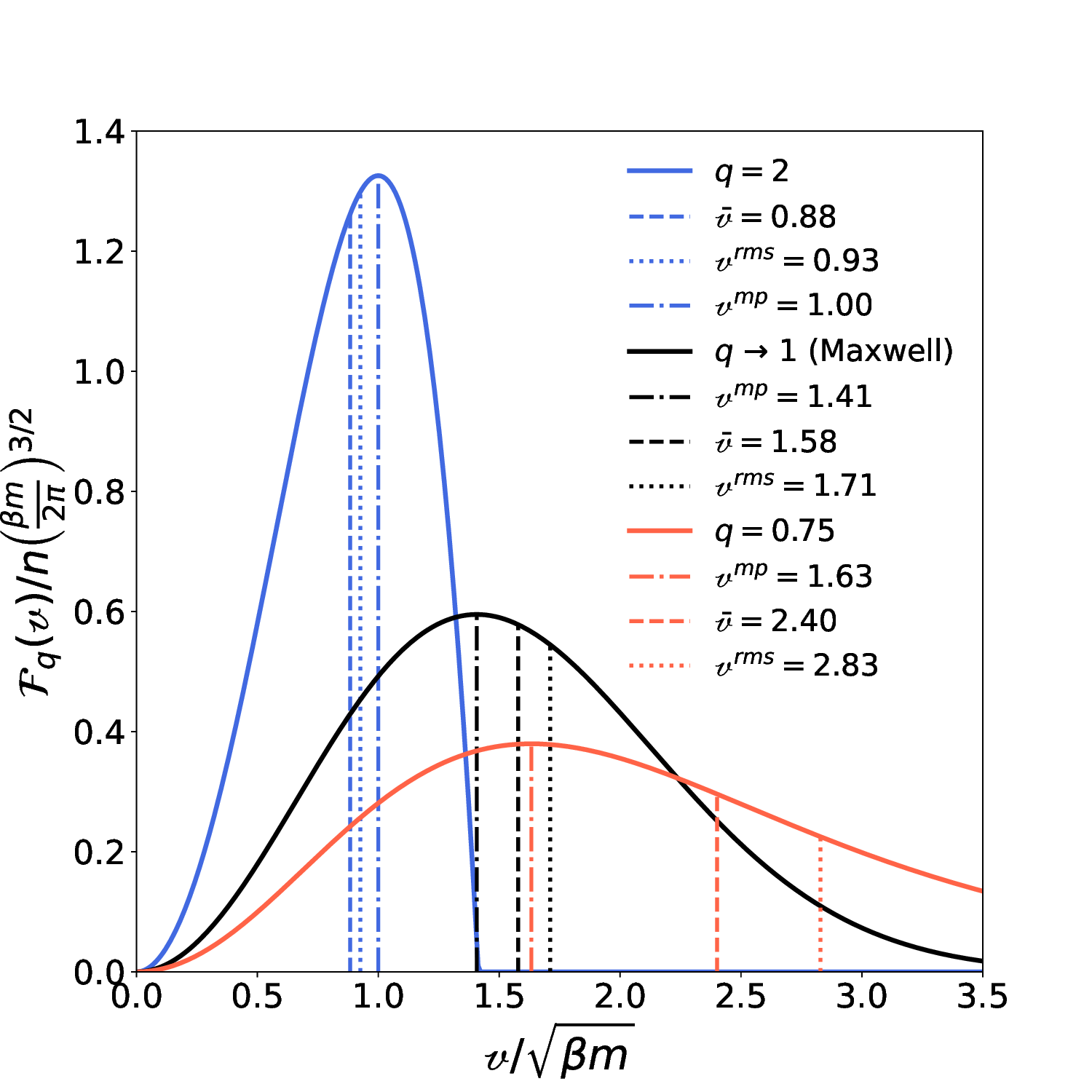}
\caption{Typical velocities for some selected values of $q$. As in figure 2, solid black line represents the Maxwellian limit ($ q \to 1 $).  The solid blue and red lines are short and fat-tailed distributions, respectively. For each color, the vertical lines are the typical velocities for the selected values of $q$ depicted in the figure. Note that in the blue case (short-tail) the sequence of typical velocities are inverted.} 
\label{f3}
\end{figure}

All corrections to the Maxwellian values are modulated only by a function that depends only on the parameter $q$. Note also that the physical quantities, except for the mean velocity, are described by the same function for $q > 1$ and $3/5 < q < 1$. Nonetheless, it is possible to show that the behavior of the Gamma functions in the mean velocity ensures a smooth and continuous transition at $q \to 1$.

In \textbf{Figure} \ref{f3} we show the dimensionless speed distribution functions (\ref{E47}) and (\ref{E64}) and their typical velocities for the selected values of $q$. We have the short-tail power-law for $q=2$ (solid blue line), the Maxwellian distribution for $q\to1$ (black), and the fat-tail power-law for $q = 0.75$ (red). The vertical lines indicate the typical velocities. When $q = 2$, we observe that $\bar{v} < v^{\text{rms}} < v^{\text{mp}}$. However, for $q<1$, the order changes to $v^{\text{mp}} < \bar{v} < v^{\text{rms}}$, which means that the typical velocities for short-tail functions do not always follow the same increasing sequence as those of the Maxwellian and fat-tail distributions. 

It is also worth noticing that there are different power-law distributions coexisting in the present literature \cite{V68,Hasegawa85,Livadiotis13,Lazar22}. In the case of fat-tailed, for instance, they have  been largely applied in the study of suprathermal space plasmas \cite{silveira21b,benetti}. Unfortunately, different from Tsallis statistics, such distributions have several formulations, and, therefore, cannot be included in the Neo-Maxwellian approach discussed here. However, the specific, isotropic, fat-tail Kappa-distribution with $\kappa > 5/2$:
\begin{equation}\label{E86}
f_{\kappa}(v) = n\left(\frac{m}{2\pi k_BT}\right)^{3/2} \frac{ \Gamma\left(\kappa\right)}{ \kappa^{3/2}\Gamma\left({\kappa} - \frac{3}{2}\right)}
\left[1 +  \frac{m v^2}{2\kappa k_BT}\right]^{-\kappa}.    
\end{equation}
is a particular case of the $d$-distribution (\ref{E63}) when one identifies $d = -1/\kappa$.  As one may show, the constraint $\kappa > 5/2$ comes from the value of average energy per particle. The two choices for $ d $ establish the well known parameter mapping $q=1-1/\kappa$ \cite{Leubner04}. 

\section{Final Remarks and Conclusion}

In order to generate equilibrium power-law distributions for a perfect gas, we have extended Maxwell's kinetic model published in 1860, by exploring the deformation of the Gaussian function, or more precisely, Euler's exponential function. The extension sought is defined by rejecting Maxwell's postulate \textbf{PI}. While the notion of isotropy of the probability density present in postulate \textbf{PII} is still satisfied, we can say that the imposition of the first Maxwellian postulate (statistical independence of the velocity distributions of each component) already fixed the Gaussian as the solution to the problem. Therefore, its rejection would naturally open the window for non-Gaussian possibilities.
 
In Section 3, we propose more general forms for ``Maxwell's postulates" \,dubbed  \textbf{PI*} and \textbf{PII*}. As a result, the Gaussian hypothesis whose origin is directly connected with the notion of statistical independence, as introduced ab initio by Maxwell, should generally be abandoned. Only in the limit $ d\to 0 $,  Maxwell's velocity distribution is recovered. As recently demonstrated \cite{LD2020,LB2025} this limit occurs when the number of particles $ N \to \infty $ (thermodynamic limit).

We also demonstrated that a generic $d$-distribution can be derived without the necessity of an effective temperature depending on the deformation parameter $d$ or even of the normalization constant. Unlike many approaches in the literature, the parameter $\beta$ retains its traditional value, $ \beta = (k_B T)^{-1} $, from the classical kinetic theory. This result is significant because it implies that we can measure the temperature of a non-Gaussian ideal gas in the same way we measure the temperature of an ideal gas in thermodynamic equilibrium, despite the correlations at low concentrations \cite{LD2020}.

We considered two cases of power-laws distributions, each corresponding to a distinct mathematical interpretation of the distribution function. For $d > 0$, the deformed function is called the short-tailed power-law, which means that it is a function limited by the Gaussian since it approaches the Gaussian case from within (see \textbf{Figure} \ref{f1}). On the other hand, for $ d < 0 $, the deformed function is called the long-tailed power-law, and its most notable feature is that it is a function outside the Gaussian. This occurs as $ d\to 0 $, with the power-law approaching the Gaussian from above. \textbf{Figures} \ref{f1} and \ref{f2} clearly illustrate this behavior for different values of $d$. Physically, both distributions for each value of $d$ are in thermal equilibrium, that is, a perfect ideal gas with a finite number of particles. 

A brief discussion on the problem of the marginal distribution involving the power-laws \cite{SPL98,beck95} is presented in Appendix \ref{C}. We observe that power laws require a more careful approach when addressing this topic, as the integral of power-laws in higher dimensions does not directly recover the functions in lower dimensions. Thus, a simple modification of the marginal distribution allows the expected results  to be recovered.

In particular, since in this context $d = \pm 2/3N$, it means that  $ q = 1 \pm2/3N$, we recover the Tsallis distribution exactly as a particular case, for which the most important physical averages have been calculated in \textbf{Table} \ref{T1}.

\vspace{1em}
\textbf{Acknowledgments:} JASL is partially supported by CNPq (310038/2019-7), CAPES (88881.068485/2014), and FAPESP (LLAMA project, 11/51676-9 and 24/02295-2). MHB is supported by FAPESP/CNPq (24/14163-3).

\section{Supplementary material}

\appendix 
\section{3-D Power-Law Normalization}\label{A}

The normalization of the 3-D function with $d>0$ can be derived from the integral
\begin{equation}\label{A1}
n = \int\int\int_{-v_{\text{max}}}^{v_{\text{max}}} B_3 \left[1 - \frac{d \beta m v^2}{2}\right]^{\frac{1}{d}} d^3 \textbf{v},
\end{equation}
to simplify the calculations, we change to spherical coordinates, such that
\begin{equation}\label{A2}
n = \int_{0}^{v_{\text{max}}} B_3 \left[1 - \frac{d \beta m v^2}{2}\right]^{\frac{1}{d}} 4\pi v^2dv,
\end{equation}
by performing the variable change
\begin{equation}\label{EA3}
u = \frac{d\beta mv^2}{2} \implies  v^2dv = \frac{1}{2}\left( \frac{2}{d\beta m}\right)^{3/2}u^{1/2}du,
\end{equation} 
and substituting into the integral (\ref{A2}), we obtain
\begin{equation}\label{A4}
n = B_3 2\pi \left(\frac{2}{d\beta m }\right)^{3/2} \displaystyle\int_{0}^{1} u^{1/2}(1-u)^{\frac{1}{d}} du.
\end{equation}

Using the \textit{Beta function} (see footnote 3)
we find that
\begin{equation}\label{A5}
n = B_3 2\pi \left(\frac{2}{\beta m }\right)^{3/2}\,\,\, \displaystyle\frac{\Gamma\left(\frac{3}{2}\right) \Gamma\left( \frac{1}{d} +1 \right)}{d^{3/2}\Gamma\left(\frac{1}{d}+\frac{5}{2}\right)},
\end{equation}
substituting $\Gamma(3/2) = \sqrt{\pi}/2$ and isolating $B_3$, we obtain
\begin{equation}\label{A6}
B_3 = n \left(\frac{\beta m }{2\pi}\right)^{3/2} \,\,\,\displaystyle\frac{d^{3/2} \Gamma\left(\frac{1}{d}+\frac{5}{2}\right)}{\Gamma\left( \frac{1}{d} +1 \right)}.
\end{equation}

In the case $d<0$ by using footnote 4 (see page 6), one obtains:
\begin{equation}\label{A7}
B_3 = n\left(\frac{ m}{2\pi k_BT}\right)^{3/2} \frac{|d|^{3/2} \Gamma\left(\frac{1}{|d|}\right)}{ \Gamma\left(\frac{1}{|d|} - \frac{3}{2}\right)}. 
\end{equation}

\section{Mean Energy and Typical Velocities ($d>0$)}\label{B}

The mean energy per particle for the $d$-distribution is given by
\begin{equation}\label{B1}
\bar{\varepsilon}_d = \frac{B_3\displaystyle\int_{0}^{v_{\text{max}}} \left[1-\frac{d\beta mv^2}{2}\right]^{\frac{1}{d}} \frac{mv^2}{2}\,\,\, 4\pi v^2 dv}{B_3\displaystyle\int_{0}^{v_{\text{max}}} \left[1-\frac{d\beta mv^2}{2}\right]^{\frac{1}{d}}4\pi v^2dv}, 
\end{equation}
writing in terms of the variable $u$,
\begin{equation}\label{B2}
\bar{\varepsilon}_d = \frac{m\pi \left(\frac{2}{d\beta m }\right)^{5/2}\displaystyle\int_{0}^{1} u^{3/2}(1-u)^{\frac{1}{d}} du}{2\pi \left(\frac{2}{d\beta m }\right)^{3/2} \displaystyle\int_{0}^{1} u^{1/2}(1-u)^{\frac{1}{d}} du}, 
\end{equation}
using the \textit{Beta function} (see footnote 3), we obtain
\begin{equation}\label{B3}
\bar{\varepsilon}_d = \frac{m\pi \left(\frac{2}{|d|\beta m }\right)^{5/2} \displaystyle\frac{\Gamma\left(\frac{5}{2}\right) \Gamma\left( \frac{1}{d} + 1\right)}{\Gamma\left(\frac{1}{d}+\frac{7}{2}\right)}}{2\pi \left(\frac{2}{d\beta m }\right)^{3/2} \displaystyle\frac{\Gamma\left(\frac{3}{2}\right) \Gamma\left( \frac{1}{d} + 1\right)}{\Gamma\left(\frac{1}{d} + \frac{5}{2}\right)}}, 
\end{equation}
using the property $z\Gamma(z) = \Gamma(z+1)$ \cite{AS1972}, equation (\ref{B3}) can be simplified to
\begin{equation}\label{B4}
\bar{\varepsilon}_d = \frac{3}{2\beta} \, \left(\frac{2}{2+5d}\right).
\end{equation}

For typical velocities, we begin with the root-mean-square velocity $v_{\text{rms}}$, given by
\begin{equation}\label{B5}
v_d^{\text{rms}}  = \left[\frac{\displaystyle\int_0^{v_{\text{max}}}  F_d(v)v^2\,\,\,4\pi v^2 dv}{\displaystyle\int_0^{v_{\text{max}}} F_d(v)4\pi v^2  dv}\right]^{1/2},
\end{equation}
the integral (\ref{B5}) is exactly the square root of the mean energy per particle (\ref{B1}) multiplied by $2/m$ [see result (\ref{B4})]:   
\begin{equation}\label{B6}
v_d^{\text{rms}} = \frac{2}{m} \bar{\varepsilon}  = \sqrt{\frac{3k_BT}{ m} \left(\frac{2}{2+5d}\right)}.
\end{equation}

The mean velocity, $\bar{v}$, is determined as follows:
\begin{equation}\label{B7}
  \bar{v}_d  = \frac{\displaystyle\int_0^{v_{\text{max}}} F_d(v)v\,\,\,\,4\pi v^2 dv}{\displaystyle\int_0^{v_{\text{max}}} F_d(v)\,4\pi v^2  dv} = \sqrt{\frac{8k_BT}{\pi m}}\frac{\Gamma\left(\frac{1}{d} + \frac{5}{2}\right)}{d^{1/2} \Gamma\left(\frac{1}{d}+ 3\right)}.  
\end{equation}

Finally, the velocity for which the function $F(v)$ is maximum defines the most probable velocity, $v_{\text{mp}}$. Thus, we need to find the condition where
\begin{equation}\label{B8}
\frac{d\left[\mathcal{F}_d(v)\right]}{dv} = 0
\end{equation}
computing the derivative and performing some calculations, we find
\begin{equation}\label{B9}
v_d^{\text{mp}} = \sqrt{\frac{2k_BT}{ m (1+d)}}.
\end{equation}

\section{The Marginal Probability Problem}\label{C}

The marginal probability, in the classical context, is defined by integrating a higher-dimensional distribution over the remaining degrees of freedom. For example, given the 3-D distribution, the 1-D function is given by:
\begin{equation}\label{E68}
f_d(v_x) = \int_{}^{} F_d(v_x,v_y,v_z)\, dv_y dv_z.
\end{equation}
In the Maxwellian case, it is straightforward to see that the marginal distribution (\ref{E68}) exactly coincides with the 1-D distribution function (\ref{E10}).

Now, let us examine what happens when the distribution function follows a power-law.

Considering $d>0$, integrating the 3-D function (\ref{E39}) over the $v_y$ and $v_z$ components, we obtain:
\begin{equation}\label{C1}
f_d(v_x) = B_3 \int_{-v_{\text{max}}}^{v_{\text{max}}} \left[1 - \frac{d m v^2}{2k_BT}\right]^{\frac{1}{d}} \,  \, dv_y dv_z,
\end{equation}
where the maximum velocity is defined as:
\begin{equation}\label{C2}
v_x^2 + v_{\text{max}}^2 = \frac{2k_BT}{d m} \Rightarrow v_{\text{max}} = \sqrt{\frac{2k_BT}{d m} - v_x^2}.
\end{equation}
We can rewrite the integral (\ref{C2}) in polar coordinates, such that:
\begin{equation}\label{C3}
f_d(v_x) = 2\pi B_3 \int_0^{v_{\text{max}}} \left[1 - \frac{d m (v_x^2 + w^2)}{2k_BT}\right]^{\frac{1}{d}} \, w \, dw,
\end{equation}
where $w^2 = v_y^2 + v_z^2$. By applying the variable change $t = (d m w^2/2k_BT)/(1 - d m v_x^2/2k_BT)$, we obtain the following integral:
\begin{equation}\label{C4}
f_d(v_x) = \frac{2\pi k_BT B_3}{d m} \left(1 - \frac{d m v_x^2}{2k_BT}\right)^{1 + \frac{1}{d}} \int_0^1 (1 - t)^{\frac{1}{d}} dt.
\end{equation}
With the aid of the \textit{Beta function} (see footnote 3) and considering $B_3$ given by equation (\ref{A6}), we arrive at the result for the marginal function:
\begin{equation}\label{C5}
f_d(v_x) = n \left(\frac{m}{2\pi k_BT}\right)^{1/2} \frac{d^{1/2} \,\, \Gamma\left(\frac{1}{d} + \frac{5}{2}\right)}{\Gamma\left(\frac{1}{d} + 2\right)} \left[1 - \frac{d m v_x^2}{2k_BT}\right]^{\frac{1}{d}+1}.
\end{equation}
The 1-D function above, determined by marginal probability, differs from the 1-D distribution function (\ref{E34}) calculated via the normalization integral in one dimension. \footnote{The integral of the power-law (3-D) over the sub-space $(y, z)$ introduces a unit increment to the exponent of the marginal distribution, as in $\int x^n dx \propto x^{n+1}$.} It is easy to verify that the marginal probability 1-D for the case $d<0$ also differs from the distribution (\ref{E35}).

In the context of the power-laws discussed here, the concept of marginal probability distributions reveals a remarkable property: the marginal distribution $F_d(v_x)$ following prescription (\ref{E68}) does not coincide with results (\ref{E34}) and (\ref{E35}) for cases $d>0$ and $d<0$, respectively.

Strictly speaking, since the $d$-distribution cannot be factorized, the shape of the 1-D distribution with the usual marginal probability depends on the number of spatial dimensions. This means that $F_d(v_x)$ is also a power-law but with an effective $d$ parameter.

In this sense, it is interesting to seek a formulation that avoids such a problem. With this concern, we propose the following modification of expression (\ref{E68}) for the $m$-dim marginal function in $n$-dimensional velocity space \cite{beck95,SPL98}:
\begin{equation}\label{E74a}
f_d(v_1,\cdot\cdot\cdot, v_m) = \int_{}^{} F_d^{\mu}(v_1, \cdot\cdot\cdot, v_n) dv_{m+1} \cdots dv_n, 
\end{equation}
where $\mu = 1 - d(n-m)/2$. For the 1-D marginal distribution (m=1) in the 3-D velocity space (n=3), equation (\ref{E74a}) becomes:
\begin{equation}\label{E74b}
f_d(v_x) = \int_{}^{} F_d^{1-d}(v_x,v_y,v_z)dv_y dv_z. 
\end{equation}
Thus, we can write the marginal function as:
\begin{equation}\label{E74c}
f_d(v_x) = B_3 \int_{-v_{\text{max}}}^{v_{\text{max}}} \left(1 - \frac{d m v^2}{2k_BT}\right)^{\frac{1}{d}-1} \,  \, dv_y dv_z.   
\end{equation}
Following the above development, it is straightforward to verify that the 1-D marginal function (\ref{E74c}) recover distributions (\ref{E34}) and (\ref{E35}) for $d>0$ and $d<0$, respectively. We leave this analysis to the reader.

\end{document}